# Bi-level Poisoning Attack Model and Countermeasure for Appliance Consumption Data of Smart Homes

Mustain Billah [1], Adnan Anwar [2,*], Ziaur Rahman [3] and Syed Md. Galib [1]

1. Department of CSE, Jashore University of Science and Technology (JUST), Jashore-7400, Bangladesh; mu.billah@just.edu.bd (M.B.); galib.cse@just.edu.bd (S.M.G.)
2. Centre for Cyber Security Research and Innovation, Deakin University-3217, Australia
3. School of Computing Technologies, RMIT University-3001, Australia; rahman.ziaur@rmit.edu.au
* Correspondence: adnan.anwar@deakin.edu.au

**Abstract:** Accurate building energy prediction is useful in various applications starting from building energy automation and management to optimal storage control. However, vulnerabilities should be considered when designing building energy prediction models, as intelligent attackers can deliberately influence the model performance using sophisticated attack models. These may consequently degrade the prediction accuracy, which may affect the efficiency and performance of the building energy management systems. In this paper, we investigate the impact of bi-level poisoning attacks on regression models of energy usage obtained from household appliances. Furthermore, an effective countermeasure against the poisoning attacks on the prediction model is proposed in this paper. Attacks and defenses are evaluated on a benchmark dataset. Experimental results show that an intelligent cyber-attacker can poison the prediction model to manipulate the decision. However, our proposed solution successfully ensures defense against such poisoning attacks effectively compared to other benchmark techniques.

## 1. Introduction

Home appliances consume a large portion of electrical energy, thus attracting researcher's attention to understand the appliance energy usage patterns in buildings [1,2]. Even appliances in standby mode can be responsible for a significant increase in electricity consumption due to their continuous consumption of low electricity demand. To determine the correlation among different factors and to assess their impact on energy management systems, prediction models can be helpful. Many applications such as estimation of energy usage patterns, energy management, load control, demand-side management and demand-side response, simulation of building performance analysis and many more will be beneficiaries of electrical energy consumption models using regression-based predictive analyses [3,4].

While these predictive models have much potential for improved energy management and efficiency, measurement information and available internet traffic communication are heavily relied upon. In reality, home and building automation systems are communication sensitive and heavily rely on information exchange. This depends on the system structure and various data access points. Any of these access points can be controlled to infiltrate a network and change load and measurement information to destabilize the framework unpredictably. Although the smart-grid paradigm has started a new era with advanced communication and control for improved reliability and efficiency, it has created new challenges as well related to cybersecurity.

In the age of the Internet of Things (IoT), cybersecurity is not only a major concern to information technology but also to critical infrastructures like energy industries [5,6]. Power and energy systems benefit from advanced probabilistic modeling and have a large impact towards improved smart-grid reliability and energy efficiency. Thus they are more vulnerable to attacks by intelligent attackers that target these computational modules. The US Department of Energy (DoE) revealed 150 effective attacks that focused on data integrity and availability attacks within power networks [7]. The European Network of Transmission System Operators for Electricity, which addresses 42 European transmission framework administrators in 35 nations, discovered proof of a fruitful digital interruption in its office network [7]. In the year 2016, cyber-attackers conducted a blackout in the Ukraine power grid for 60 min [8]. Intruders gained access to the operation center of a western Ukrainian power company in 2015 and disconnected power from 225,000 families. A US report reasoned that the attack was launched through 'spear phishing' emails [9]. A detailed investigation of cyber-attack threat models is presented in [10] while the countermeasures are summarized in [11].

Accurate predictions can help decision makers to determine patterns of future consumption to better plan and optimize energy consumption, which will reduce power loss and increase energy efficiency of the energy systems. Similarly, building energy prediction will help to improve the energy efficiency of the building energy management system. However, corrupted prediction can lead to safety hazards, damage of power system assets and financial losses. For example, in poisoning attacks, intruders try to manipulate the training dataset to influence the outcome of a predictive model [12]. In these types of attacks, the attacker corrupts the learning model such that the model fails to correctly predict on new data during the testing phase. Thus, indirect access to the model is established, which can be used by the attacker to handle the model even in future. In another case, misclassification of unseen data may happen in evasion attacks during the testing phase. Here, an intruder may bypass a data security mechanism to convey an attack or other type of malware to a target model [13]. In privacy attacks, private information in training data can be stolen and utilized to perceive the private behavior of people [14].

However, among many cybersecurity issues, data poisoning attacks pose a great threat to energy consumption prediction. These types of attacks have been analyzed in many application domains such as worm signature generation [15], denial-of-service attack detection [16], PDF malware classification [17], etc. Although attacks on machine learning models, known as adverserial machine learning, has been an emerging research area, not many research works have been reported on manipulating machine learning models of energy consumption data. Such models should be updated regularly because data are generated continuously from different energy sources. In such cases, attacks become easier to mount on the models not only in the machine learning modules but also during data transfer. For instance, in a national energy/load management system, data are collected from different regions of a country through an online system, in which intruders can also put forward data of their choice by generating a data integrity attack. By controlling a couple of devices, intruders can submit counterfeit data, which are then utilized for preparing models applied to an enormous set of choices. Again, it is a challenging task to prevent poisoning attacks as current techniques against adversarial poisoned data perform poorly. In this work, a bi-level poisoning attack is carried out against an aggregated energy usage prediction model considering three popular regression models: (a) the ordinary least squares (OLS) model, (b) the ridge regression model and (c) the lasso regression model. Analyses are carried out based on a real-world dataset that contains different data sources and environmental parameters. Regression is extensively used for prediction purposes in many domains where a model tries to minimize a loss function and predict a numerical variable based on some predictor variables.

This paper aims to address the vulnerability due to a poisoning attack on household energy data and deal with potential countermeasures. Here, we consider bi-level poisoning attacks, where attackers inject poisoned data in two steps: during data propagation from devices of the smart home to the control center (during data transit) and during training of the machine learning model. Specifically, the contributions can be summarized as follows:

1. We assess a bi-level data poisoning strategy based on a sparse attack vector and optimization-based attack, which successfully corrupts the energy prediction model of home appliances (See Section 3);
2. An effective solution for the poisoned energy prediction model is also implemented. The proposed defense strategy is evaluated on various benchmark regression models (See Section 3);
3. Apparently, to the best of our knowledge, this is one of the earliest works on the attack and defense of poisoning attacks on 'household energy prediction models'. Proposed methods are tested on a benchmark dataset from the UCI data repository (Section 5).

## 2. Related Work

This article deals with the cybersecurity vulnerabilities and countermeasures of the predictive models for household energy usage. Hence, we have divided the review into two broad categories.

### 2.1. Related Work on the Household Prediction Models

Diverse information and techniques are utilized in the existing literature to comprehend forecast models of energy utilization of home appliances. A methodology to appraise building energy utilization is assessed from the standard datasets in [18]. Hourly energy consumption is predicted from service bills using predetermined coefficients. Various residential accessory loads, including a fridge, PC, TV and clothes washer, are modeled using the Norton equivalent technique in [19]. Day-by-day energy use profiles for significant home devices are investigated in [20], which claims that refrigerators show better uniformity than user-dependent accessories such as clothes washers. The authors in [21] develop a model that can detect and estimate different home appliance loads using an explicit duration hidden Markov model. The review paper in [22] finds crucial appliances and their parameters affecting electricity consumption in domestic buildings. Another study focuses on electrical appliance modeling for accurate energy simulations for buildings [23]. However, most of the models presented here are concerned with building simulation studies. However, a number of studies are also found in the literature dealing with electricity load prediction in operational phases.

A large variety of parameters are considered in the prediction models of electricity demand. The most important parameters for such models are rainfall index, time of day, outdoor temperature, global solar radiation, wind speed, etc. [24]. The impact of climate-related factors on monthly power demand are studied in [25]. Another study [26] considers brief-stretch power estimation for different houses and finds temporal distribution as a significant factor. The investigation in [27] uncovers that climate, area, and floor zone are the main factors, and the numbers of fridges and entertainment devices are the most impacting determinants, of day-by-day least utilization. To predict an individual appliance's energy consumption, a system was developed by [28] where different data, for example, past utilization, season, month, etc., were utilized to supervise the model.

### 2.2. Related Work on the Security Vulnerabilities of the Predictive Models

It is important to note that prediction models have a large probability of security vulnerabilities to potential attackers. Recently, the safety aspects of machine learning models have attracted the attention of researchers. A large variety of attacks on different domains have been designed and analyzed, which are especially in three categories: poisoning attacks, evasion attacks and privacy attacks. Such attacks have been practically demonstrated in many application domains.

A family of poisoning attacks has been investigated against a support vector machine (SVM)-based handwritten digit recognition system in [12]. Such attack infuses uniquely created training data that expand the SVM's test data. Another work on malware detection in [17] reveals that feature selection techniques can be seriously compromised under poisoning attacks. A poisoning attack and a corresponding defense mechanism are proposed in [16], with regard to a specific anomaly detector for identifying peculiarities in background networks. A noise injection attack on a syntactic-based programmed worm signature generator is introduced in [15].

A group of evasion attacks is constructed in [29] to evaluate the robustness of a neural network. Three standard datasets (MNIST, CIFAR-10, ImageNet) are used to evaluate proposed attacks. In the paper [30], the authors tentatively research the viability of classifier evasion attack utilizing a genuine, established framework called PDFRATE. Another work [31] shows how an attacker can abuse AI, as utilized in the SpamBayes spam filter, to render it pointless regardless of whether the attacker's entrance is restricted to just 1% of the training messages. In the work [13], the authors present a basic yet compelling gradient-based methodology that can be abused to deliberately evaluate the security of a few, generally utilized classification tasks against evasion assaults. The methodology is assessed on the security task of malware recognition in PDF files, and shows that such frameworks can be effortlessly sidestepped.

A large amount of research can be found on privacy attacks in the literature. Privacy attacks in pharmacogenetics are investigated in [32], wherein AI models are utilized to control clinical medicines dependent on a patient's background and genotype. It justifies that, given the model and some segment data about a patient, attackers can foresee the patient's hereditary markers. Ref. [33] quantitatively explores how AI models spill data about the individual records on which they were trained. Utilizing realistic datasets, including a clinic release dataset whose membership is sensitive from the data protection point of view, they show that these models can be defenseless against membership inference attacks.

While a number of research works have focused on household energy consumption prediction [18–28], the security issues of these models are not well addressed in the literature. However, though there are many works on poisoning attacks in machine learning models for different domains [12,13,15–17,29–33], the impact of a false data injection attack during communication along with a poisoning attack on machine learning models specifically for energy systems (to the best of our knowledge) cannot be found in the literature. In this paper, we perform bi-level poisoning attacks on regression models for predicting the energy usage of household appliances.

## 3. Proposed Bi-level-Poisoning-Based Adversarial Model on Energy Data

A household prediction model heavily relies on the sensing and decision making as shown in Figure 1. Information is sensed from smart home appliances and sent to a server through an advanced communication technique where decision making occurs. However, an intelligent attacker can perform one specific type of attack known as a data integrity attack in two different ways. One is during the communication of the information that they can manipulate, which is similar to the man-in-the-middle attack or a spoofing attack. Another involves poisoning the machine learning model. Based on this framework and model, we have proposed a bi-level poisoning attack model.

### 3.1. Poisoning Attack During Communication

Different devices in the smart home send signals and streaming data periodically to the central machine learning module. An attacker may capture and possibly alter the data between smart home devices and the server. The attacker may also be able to intercept all relevant signals passing between them and corrupt a fraction of the entire dataset. In this section, we construct a false data injection attack during communication from smart home appliances to the server.

False information infusion assaults require the attackers to know the current design of the smart home framework. In the event that the attacker can determine the current arrangement of the automated smart building management system, attackers can inject poisonous measurements that will mislead the decision making process of the machine learning module. As this system configuration does not change frequently in a smart home, it is trivial for the assailants to acquire such design data to launch these attacks. Another requirement for the attackers is the manipulation of the sensor measurements. The attackers need to truly mess with the sensor, or manipulate the sensor measurements before they are used for training in the machine learning module. Strong protection against unauthorized physical access to these sensors will make it nontrivial to control the sensor estimations. Be that as it may, it is the beginning stage of our exploration, and the current outcomes can act as a basis for future examination of more complicated models. We consider the scenario where the assailant is obliged to get to some particular sensors and perform arbitrary bogus information infusion assaults, in which the assailant expects to discover any assault vector as long as it can prompt an off-base assessment of state factors.

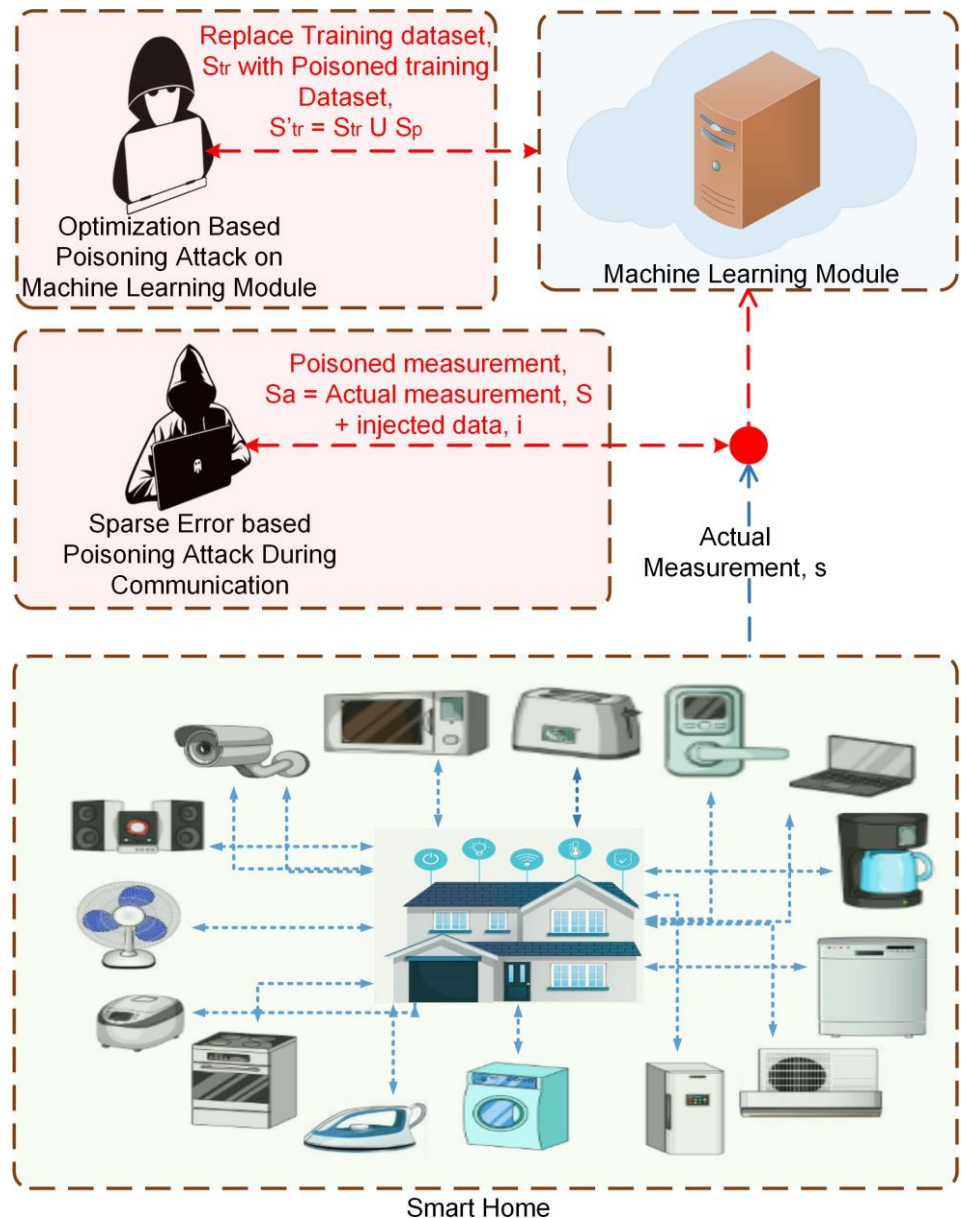

**Figure 1.** Proposed system architecture. Energy consumption data for each home appliance are sent over a wireless network to the central machine learning server (ML module). During communication from smart home to ML module, attacker may inject false data. However, during training of the ML models, training data can be manipulated by attackers.

We expect that there are $p$ sensors that give p estimations $(s_1, \ldots, s_p)$ and these sensors send a total of $q$ observations $(o_1, \ldots, o_q)$ periodically. The relationship between these sensor measurements and $q$ observations can be characterized by a $q \not{p}$ matrix $M$.

Allow $s_a$ to address the vector of noticed estimations that may contain pernicious information. $s_a$ can be addressed as

$$s_a = s + i \quad (1)$$

where $s = (s_1, \ldots, s_p)$ is the vector of original measurements and $i = (i_1, \ldots, i_p)$ is the pernicious information added to the first estimations. We allude to $i$ as an assault vector. The $k$th component $i_k$ being nonzero implies that the assailant bargains the $k$th sensor and afterward replaces its unique estimation $s_a$ with a fake estimation $s_k + i_k$. The attacker can pick any nonzero self-assertive vector as the assault vector $i$ and afterward build the pernicious estimations $s_i = s + i$.

We accept that the attacker approaches $x$ explicit sensors. Instinctively, the assailant can just adjust the estimations of these $x$ sensors. Therefore, the assailant cannot basically pick any $v = [v_1, ..., v_p]^T$ and use $i = Mv$ as the assault vector. For those sensors that cannot be accessed by the attacker, the infused errors should stay 0.

Formally, we let $L_{sensors} = l_1, \ldots, l_x$ be the arrangement of lists of the $x$ meters that the aggressor approaches. The aggressor can adjust estimations $s_{l_z}$, where $l_z \epsilon L_{sensors}$. To dispatch a bogus information infusion assault without being distinguished, the aggressor needs to discover a nonzero assault vector i = $(i_1, \ldots, i_p)$ such that $i_l = 0$ for $l\ /\ L_{sensor}$ and i is a direct mix of the segment vectors of $M$ (i.e., $i = Mv$).

As we consider an arbitrary bogus information infusion assault that causes incorrect estimation of the observations, the errors that infuse into some unacceptable assessment could be useful.

Consequently, the assault vector $i$ satisfies the condition:

$$i = (i_1, ..., i_p) = Mv \quad (2)$$

with $i_l$ = 0 for $l\ /\ L_{sensor}$, where $L_{sensor}$ is the arrangement of files of the meters that can be accessed by the aggressor.

### 3.2. Poisoning Attack on the Predictive ML Module

The second level of attack performs a poisoning attack on a linear regression model. We consider the optimization-based poisoning attack proposed by [34]. In an ideal case, a linear regression model generates a function $f(x, \theta) = w^T x + b$ after the training stage. This linear function predicts the estimation of $y$ at $x$. A regularization parameter preserves the generalization capability of the algorithm on unseen data. Based on the types of regularization term, different linear regression methods are used: ordinary least squares (OLS), ridge regression, LASSO etc. Information, $s$, from the smart home appliances is sent through an advanced communication system to the central server (machine learning module), which is corrupted by the man-in-the-middle attack. Thus, poisoned data, $s_a$, are stored in the server that is used for training the ML module. Let us assume the data stored in the server are denoted by $S = s_a$.

A proposed second-level poisoning attack tries to corrupt the learning model during the training phase. This corrupted model modifies the prediction results on new data in the testing phase. Both the white-box and black-box attacks are considered in this attack. In a white-box attack, the intruder has knowledge about the training data $S_{tr}$, list of features $x$, learning algorithm $LA$ and prepared boundary parameters $\gamma$. For white-box attacks, this optimization problem can be written as:

$$arg\ max_{S_p}\ \ \ LF(S', \gamma^p_*) \quad (3)$$

where $\gamma^p_* \in arg\ min_\gamma\ L(S_{tr}\ U\ S_p, \gamma)$ and $LF$ is the loss function; $S'$ is the untainted dataset.

However, in a black-box attack, it has no information on the training set $S_{tr}$ yet can gather a substitute informational collection $S'_{tr}$. The list of features $x$ and learning algorithm $LA$ are known, while the prepared boundary parameters $γ$ are unknown. Optimization of $LA$ on the substitute data set $S'_{tr}$ can be helpful in estimating these parameters. The attack's capability in the optimization-based attack is usually confined to the upper-bounding of the number $p$ of poisoning points that can be infused into the training information. The feature values and response variables are randomly selected within a range of [0, 1]. However, poisoning rates less than 25% are well advised, because normally the attacker can handle just a small segment of the training information. More details on the optimization-based poisoning attack (Algorithm 1) can be obtained from [34].

---

**Algorithm 1:** Poisoning Attack on the Predictive ML Module.

**Result:** The final poisoning attack samples $S_p \leftarrow S_p^{(i)}$
**Initialization:**
$i \leftarrow 0$ (iteration counter)
$γ^{(i)} \leftarrow arg\ min_γ\ LA(S \cup S_p^{(i)}, γ)$
**while** $|LF^{(i)} - LF^{(i-1)}| < e$ **do**
    $LF^{(i)} \leftarrow LF(D', γ^{(i)})$
    $γ^{(i+1)} \leftarrow γ^{(i)}$
    **while** $c<=p$ **do**
        $x_c^{(i+1)} \leftarrow line\_search(x_c^{(i)}, \nabla_{x_c} LF(D', γ^{(i+1)}))$
        $γ^{(i+1)} \leftarrow arg\ min_γ\ LA(S \cup S_p^{(i+1)}, γ)$
        $LF^{(i+1)} \leftarrow LF(S', γ^{(i+1)})$
    **end**
    $i \leftarrow i + 1$
**end**

---

## 4. Proposed Defense Mechanism against Bi-Level-Poisoning-Based Adversarial Model

Proposed bi-level poisoning attack deploys two levels of attack: one is during communication from smart home appliances to the ML module, another is while training the ML module. Hence, we have proposed two levels of security mechanism with two distinct algorithms, which will work collaboratively and provide robust defense against such kinds of poisoning attacks.

### 4.1. Defense Mechanism Against Poisoning Attack During Communication

The first level of poisoning attack manipulates measurement sensors. As discussed earlier, from a set of measurement sensors with some observations a 2-D matrix will eventually form. Hence, this kind of attack is similar to alteration of a 2-D matrix where the attack vector represents a sparse data set. Defending this attack is similar to recovering a low-rank matrix with a small corrupted portion. If it is viewed from the point of a robust classical principal component analysis (PCA) problem, a large number of application domains are found to face such a problem. Principal component analysis (PCA) assumes that high-dimensional information lies almost in a much lower-dimensional subspace. The correct assumption of this subspace is important to reduce the dimension of the data and to process, analyze, compress and visualize the data. If $s_a = s + i$, where $s_a$ is a large matrix of data arranged as columns, s is a rank-r matrix and i is a matrix of Gaussian random variables, the goal of PCA is to estimate s.

Though PCA finds the optimal estimate of the subspace, it breaks down even under small corruptions. Robust PCA (RPCA) [35] can recover the low-rank matrix s from $s_a$ with gross but sparse errors i. However, although it provides superior convergence rates, the complexity of robust PCA is too high. An accelerated proximal gradient algorithm is faster and more scalable for estimating s from $s_a$ in the presence of sparse error i. To defend the first-level poisoning attack on the dataset, we apply nuclear norm minimization and a 1-norm-minimization-based convex programming surrogate named the accelerated proximal gradient algorithm (Algorithm 2) [36].

The problem of the robust PCA can be formulated as follows:

$$p^* = min_{s,i} \ ||s||_* + \lambda ||i||_1$$
$$s_a = s + i \tag{4}$$

A relaxation of Equation 4 is considered in [36] as follows:

$$min_{s,i} \ \mu ||s||_* + \mu\lambda ||i||_1 + \frac{1}{2}||s + i - s_a||_F^2 \tag{5}$$

Instead of fixing $\mu$ to any small value, convergence can be achieved in the accelerated proximal gradient algorithm by repeatedly decreasing the value of $\mu$.

---

**Algorithm 2:** Defense Mechanism Using Accelerated Proximal Gradient Algorithm in First-Level Attack.

**Result:** $s = s_k, i = i_k$
**Input:** Observed Matrix $s_a$, parameter $\lambda$
**Initialization:**
$k = 0, s_0 = s_{-1} = 0, i_0 = i_{-1} = 0, r_0 = r_{-1} = 0, \bar{\mu} > 0; \eta < 1;$
**while** *not converged* **do**

$\quad Y_k^s = s_k + \frac{r_{k-1}-1}{t_k}(s_k - s_{k-1}), Y_k^i = i_k + \frac{r_{k-1}-1}{t_k}(i_k - i_{k-1});$
$\quad G_k^s = Y_k^s - \frac{1}{2}(Y_k^s + Y_k^i + s_a);$
$\quad (U, \Sigma, V) = svd(G_k^s), s_{k+1} = U i_{\mu k}/2[\Sigma]V^T;$
$\quad G_k^i = Y_k^i - \frac{1}{2}(Y_k^s + Y_k^i + s_a);$
$\quad i_{k+1} = i_{\lambda\mu k}/2[G_k^i];$
$\quad r_{k+1} = \frac{1 + \sqrt{4\eta_k^2+1}}{2};$
$\quad \mu_{k+1} = max(\eta\mu_k, \bar{\mu});$
$\quad k \leftarrow k+1$

**end**

---

*4.2. Defense Mechanism Against Poisoning Attack on the Predictive ML Module*

The second-level attack is on the linear regression model that poisons the training dataset so that the prediction model gives the wrong decision while testing. Existing protection propositions can be grouped into two classes: noise versatile safeguards and adversarially versatile safeguards. The main idea behind these noise-resilient regression algorithms is to distinguish and eliminate anomalies from a dataset. While these techniques ensure robustness against commotion and anomalies, an attacker can in any case produce harmful information that influences the prepared model. Specifically, an assailant can produce poisoning points that are basically the same as the valid information dissemination (called inliers); however, these can in any case deceive the model. Incidentally, these current regression techniques are not strong against inlier attack points picked to maximally deceive the assessed regression model.

However, recently proposed adversarially-tough regression algorithms regularly give guarantees under solid presumptions about information and clamor dissemination. These algorithms are based on some assumptions such as information, commotion fulfilling the sub-Gaussian assumption, the component matrix having a low rank and it being possible to project the matrix to a lower dimensional space. Each of these strategies have provable robustness guarantees; however, the suppositions on which they depend are not typically fulfilled.

To defend such an attack, the TRIM defense algorithm [34] is used, which takes a principled approach instead of just removing outliers from the training set. It applies an iterative approach and in each iteration, it estimates the regression parameters, $\gamma$, and trains on a subset of points with the lowest residuals. Moreover, a loss function $LF$ is also used, calculated iteratively on a disparate subset of the residuals. The size of the original training set $s_{tr}$ is n and the attacker injects poisoned samples $S_p$, where $p = \beta.n$. We have to ensure that $\beta$ is less than 1 so that the majority of training data remain unpoisoned. The linear regression model can be trained with a subset of legitimate training points of size $n$, if all $p$ poisoning points can be identified. However, separating the legitimate and attacked points is difficult as the true distribution of the unpoisoned training data is clearly unknown. The TRIM defense mechanism tries to determine a subset of training points that have the lowest residuals. In this work, we applied trimmed optimization techniques (Algorithm 3) for the adversarial linear regression model, which has been effectively assessed for the energy dataset.

---

**Algorithm 3:** Defense Mechanism Against Poisoning Attack on the Predictive ML Module.

**Result:** Trained parameters $\gamma$
**Input:** Training Data, $S = S_{tr} \cup S_p$, attack points, $p = \beta.n$
**Initialization:**
$i \leftarrow 0$ (iteration counter)
$RS^{(0)} \leftarrow$ arbitrary subset of size n of $\{1...N\}$ $\gamma^{(0)} \leftarrow arg\ min_\gamma\ LA(RS_{(0)}, \gamma)$
**while** $(i > 1 \wedge CL^{(i)} = CL^{(i-1)})$ **do**
  $RS^{(i)} \leftarrow$ subset of size n that min. $LA(S^{RS^{(i)}}, \gamma^{(i-1)})$;
  $\gamma^{(i)} \leftarrow arg\ min_\gamma\ LA(S^{RS^{(i)}}, \gamma)$;
  $CL^{(i)} = LA(S^{RS^{(i)}}, \gamma^{(i)})$;
  $i \leftarrow i + 1$;
**end**
**return** $\gamma^{(i)}$

---

## 5. Results and Discussion

We carried out our attack and protection mechanisms in Python, utilizing the numpy and sklearn bundles. We utilized the standard cross-validation strategy to divide the whole dataset into training, testing and validation sets. We utilized two primary measurements for assessing our calculations: mean square error (MSE) for the adequacy of the assaults and protections, and running time for their expense.

### 5.1. Datasets

The dataset we used in this work comes from the UCI AI storehouse [37]. It contains the temperature and dampness of various rooms in a low-energy house from a remote sensor network, information from a climate station and one sub-metered electrical fuel utilization source (lights). The low-energy house followed passive house certification design rules, thus having a yearly heating and cooling load of about 15 kWh/m$^2$ each year. The data were gathered each 10 min from various electric energy meters including a heat recuperation ventilation unit, homegrown high-temperature water heat siphon and electric baseboard radiators. Another sub-metered load (lights) was used for the investigation since it has been demonstrated to be a decent indicator of room inhabitance when joined with relative stickiness estimations. Weather information from the closest air terminal climate station was converged by date and time in this investigation to assess its effect on the forecast of the energy utilization of appliances.

### 5.2. Effects of Poisoning Attacks on Energy Consumption Data

In this section, we analyze the effects of sparse-error-based and optimization-based bi-level poisoning attacks on energy datasets for three popular regression models: ordinary least squares (OlS), LASSO and ridge regression. Figure 2a shows the mean square error (MSE) of the proposed bi-level poisoning attack for a 5% poisoning rate. Overall, the ordinary least squares (OLS) model shows the highest error and the ridge model shows the lowest error of 0.07 and 0.04, respectively. For the attacks without poisoning, all the models show similar errors, although OLS has less error than the other models. Figure 2b shows the mean square error (MSE) of the proposed bi-level poisoning attack for a 10% poisoning rate. It is seen that the proposed bi-level poisoning attack affects the OLS model most, whereas the ridge model is least affected. The lasso model shows a significant MSE value of 0.07 compared to a 5% poisoning error.

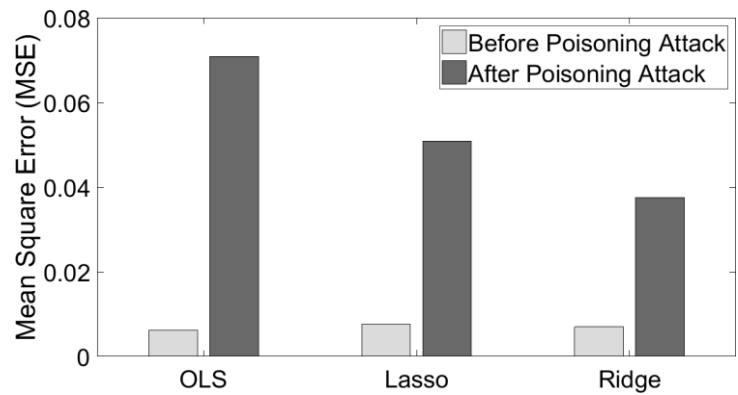

**(a)** 5% poisoning effect on three linear regression models

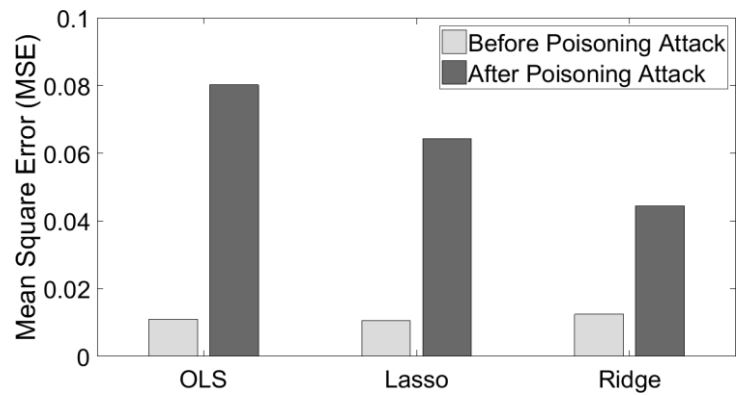

**(b)** 10% poisoning effect on three linear regression models

**Figure 2.** Mean square error (MSE) of attacks on different regression models.

Effects of bi-level poisoning attacks on different models for various poisoning rates can be well-understood from Figure 3. Various poisoning rates starting from 5% to 25% are assessed for three models. The ordinary least squares model for energy consumption prediction is mostly affected by a data poisoning attack with varying MSE for different poisoning rates. A change in poisoning rate from 10% to 15% results in a sharp change in error for the OLS model. Though it is assumed that an increase in poisoning rate will increase MSE, for the case of OLS, MSE increased up to the 15% poisoning rate, but after this point, there was a fall in MSE to 20%, from which point it started to increase again. These behaviors of OLS can be revealed as we focus on the working principles of the OLS method. The ordinary least squares (OLS) linear regression model is not penalized for its selection of weights. During the training stage, the model may place a large weight on the features that seem to be important. As a large amount of predictor variables are affected by poisoning attacks and such manipulation can make these variables correlated, OLS parameter estimates face a large variance that makes the model unreliable.

However, both the lasso and ridge regression models show exponential curves for increasing poisoning rates. The ridge model showed a gradual rise in MSE as the poisoning rate increased from 5% to 25%. The lasso model faced an irregular increase in MSE as poisoning rate increased. Unlike OLS, the lasso model is penalized for the sum of absolute values of the weights. Therefore, the absolute values of weight will not only be reduced but also many will tend to be zeros. However, ridge penalizes the model for the sum of squared values of the weights. Here, the weights have smaller absolute values and tend to penalize the extremes of the weights, thus weights are more evenly distributed. The proposed bi-level poisoning attack injects poisonous data in such a way that a lot of predictor variables are manipulated. The relevancy of these affected features to the prediction results in different effects on the lasso and ridge models. For the case of the lasso model, relevancy with all predictor variables generates more errors in prediction tasks, whereas the ridge model faces fewer errors than the lasso model due to strong relevancy with all the feature variables.

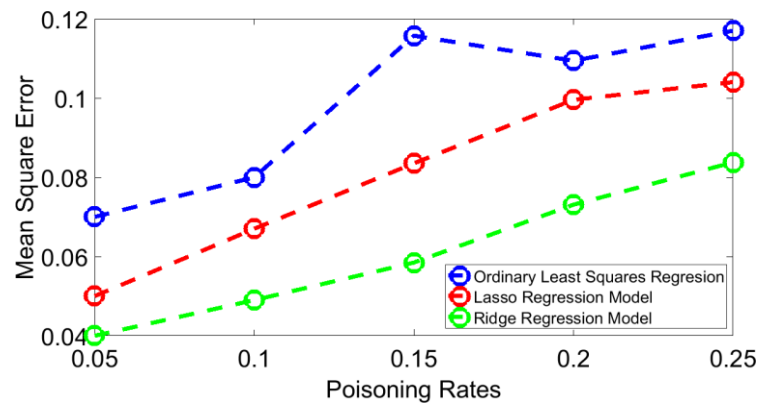

**Figure 3.** MSE for different poisoning rates.

Table 1 shows the real effect of proposed bi-level poisoning attacks on the energy dataset. MSE results obtained with the proposed attack were translated into application-specific parameters. In the prediction model of electrical energy consumption, the goal is to predict probable energy use (in watt-hour (WH)) for home appliances. Actual energy consumption at any time instant is 580 WH, which is predicted as 579.96% by the model without poisoning. However, every regression model is conquerable by poisoning attacks with a significant change by a factor of 2.87 or more for a 10% poisoning rate. The same scenarios are also found in the case of smaller poisoning rates. For a 5% poisoning rate, the change in energy consumption is 128% for the ridge model, 139% for OLS and 145% for lasso regression.

**Table 1.** Initial energy consumption (WH) and prediction (WH) after poisoning attack for different poisoning rates.

| Poisoning Rate | Appliance's Energy Use (in WH) | Prediction without Poisoning | Change in Predicted Value | | |
|---|---|---|---|---|---|
| | | | OLS | Ridge | Lasso |
| 1% | | | 44.07% | 23.25% | 27.46% |
| 2% | | | 91.10% | 53.11% | 51.62% |
| 3% | | | 70.61% | 70.64% | 103.65% |
| 4% | 580 | 579.96 | 95.17% | 105.28% | 118.17% |
| 5% | | | 139.25% | 128.53% | 145.64% |
| 6% | | | 186.28% | 158.39% | 169.79% |
| 7% | | | 165.79% | 175.92% | 221.83% |
| 8% | | | 190.35% | 210.56% | 236.34% |
| 9% | | | 234.43% | 233.82% | 263.81% |
| 10% | | | 281.46% | 263.67% | 287.96% |

The attack additionally needs to meet a time prerequisite to expand its probability of staying stealthy. If the time needed for attack development is too high, the working conditions may change and this will affect the likelihood of the assault being detected. Henceforth, in this part we assess the performance of the proposed bi-level poisoning attack in terms of effectiveness, e.g., how quickly it can create an assault. We will compare the performance sparse-error-based attack against the optimization-based attack. The time requirements for both attack constructions for two different poisoning rates are plotted in Figure 4.

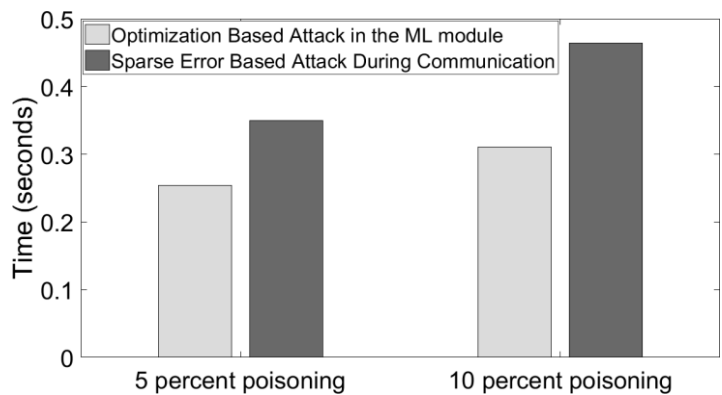

**Figure 4.** Elapsed time (in seconds) for attack construction for optimization-based attack and sparse-error-based attack.

Overall, poisoning more data points requires more time for both types of attack. In addition, the sparse-error-based attack during communication from smart home appliances to the ML module requires more time than the optimization-based attack in the ML module. For instance, it requires 0.35 s for the sparse-error-based attack while it needs only 0.25 s for the optimization-based attack for a 5% poisoning rate. However, a significant increase in time requirements is noted for the sparse-error-based attack than the optimization-based attack for a 10% poisoning rate. Therefore, the optimization-based attack construction method requires the least computational time.

*5.3. Defense Algorithms*

In this part, we assess the accelerated proximal gradient algorithm (APG) and TRIM defense mechanism separately against the bi-level poisoning attack. Figure 5a–c show MSEs for OLS, LASSO and ridge regression, respectively, for the model with no defense mechanism and the model with the TRIM defense algorithm.

The optimization-based attack generates inlier points with comparative conveyance like the training dataset, for which the TRIM technique is much more effective. For OLS, LASSO and ridge regression, the mean square error (MSE) of the TRIM defense algorithm is within 1% of the original models. This demonstrates that the TRIM technique is a significant defense mechanism against a poisoning attack in the ML module. In addition, the defense we evaluated ran very fast, taking an average time of only 0.02 s.

However, the sparse errors introduced during the first-level attack are minimized using the accelerated proximal gradient (APG) algorithm (Figure 6). To inject sparse errors, the intruder should have sound knowledge on the running configuration of a smart-home system. The non-changing behavior of the architecture of smart home systems adds an extra benefit to the attacker. However, it is assumed that only a few sensors in a smart home can be accessed by the attacker. In Figure 6, only 50 observations are illustrated, where among 28 sensors and data sources, only the temperature sensors are assumed to be accessed by the intruder. We apply poisonous data in the temperature sensors. As the APG security mechanism is deployed, it successfully captures the poisonous data $s_a$ and separates the original measurement $s$ and injected sparse error $i$.

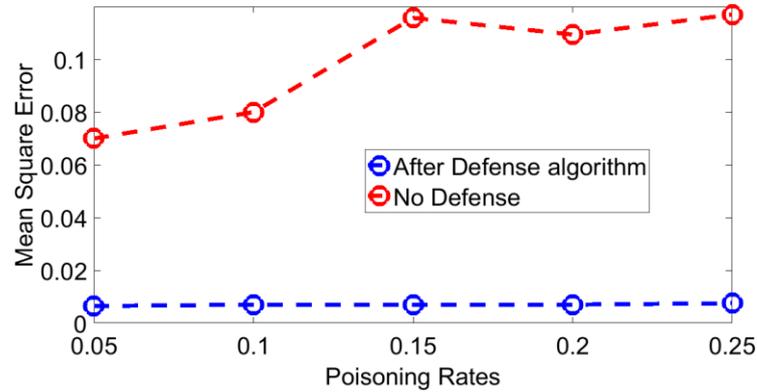

**(a)** Effect of defense mechanism on ordinary least squares model

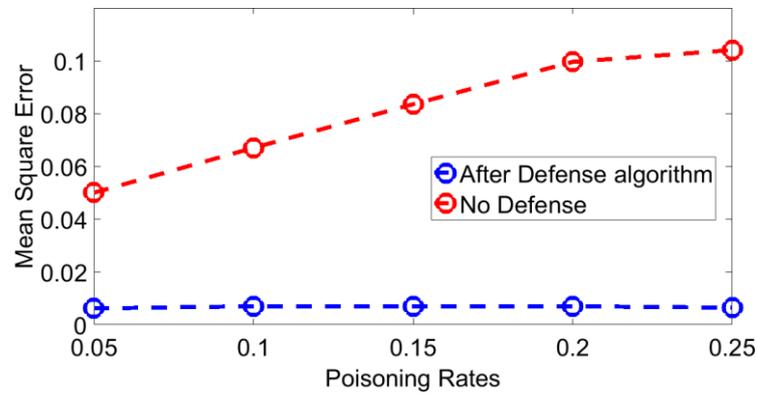

**(b)** Effect of defense mechanism on lasso model

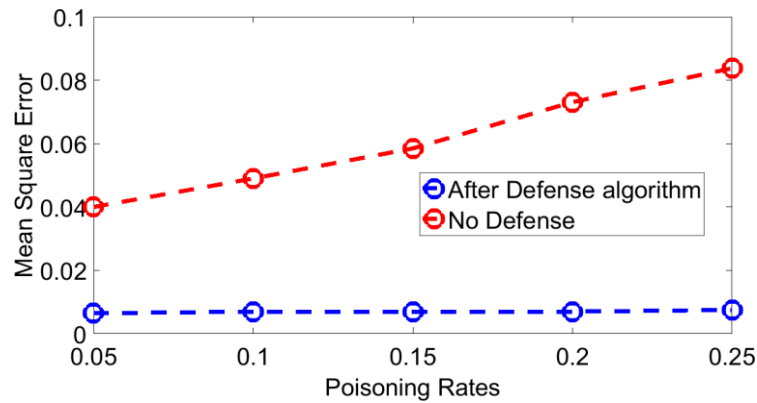

**(c)** Effect of defense mechanism on ridge model

**Figure 5.** Effects of TRIM defense mechanism on different regression models are expressed by mean square error (MSE). MSE without defense mechanism and after applying TRIM defense are compared in this figure.

However, no defense mechanism is one hundred percent capable of detecting and recovering poisonous data. Especially for APG (Figure 7), it is seen that although it reduced the error rate to a great extent, an increasing graph in error rate is also observed in accordance with an increase in poisoning rate. For example, in the case of 25% poisoning, more than 0.04% error exists in spite of the deployment of the APG algorithm.

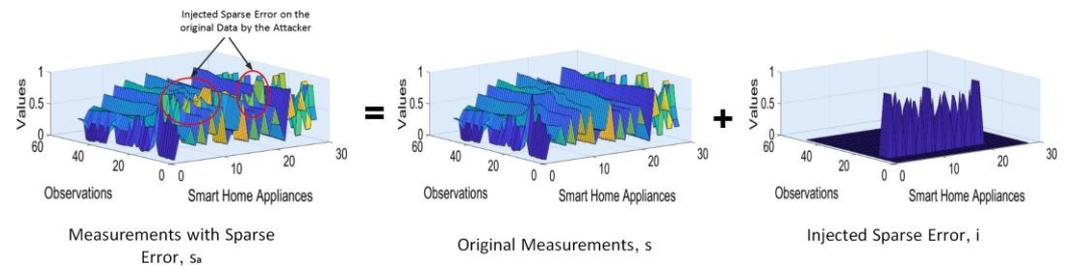

**Figure 6.** Accelerated proximal gradient algorithm for handling sparse-error-based attacks during communication from smart-home appliances to the ML module.

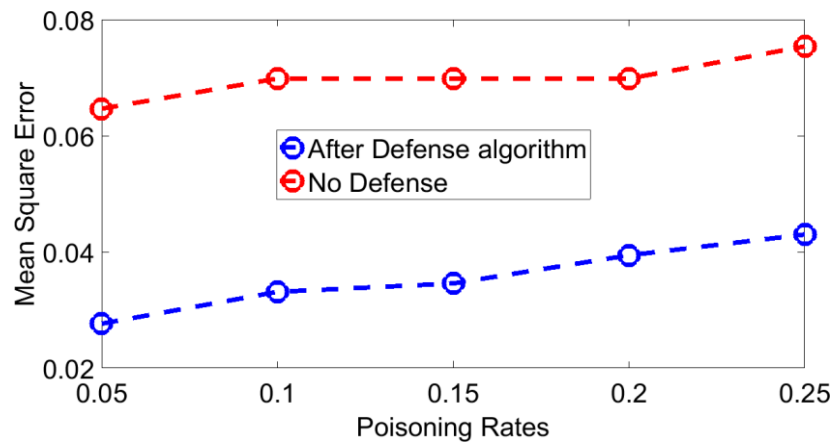

**Figure 7.** Effect of accelerated proximal gradient algorithm.

## 6. Conclusions

In this paper, bi-level poisoning attacks on linear regression models for forecasting energy utilization of appliances in a smart home were performed. These reveal that poisoning attacks have an adverse effect on building energy consumption prediction models. In particular, bi-level poisoning attacks during communication and training of the ML module may worsen the situation. Even if the attacker can poison at a rate of 10%, it can change the predicted value up to 287% in the proposed model. The prediction will become worse if the poisoning rate is increased. To tackle this emerging challenge, a combined multiple linear regression attacking solution was proposed against the poisoning attacks on the prediction model. This combined defense mechanism is time efficient and can reduce the mean square error to a great extent. Attacks and defense algorithms were systematically evaluated on a standard data set containing data from a wireless network, climate information from an air terminal station and energy consumption data from home appliances. The real implications of poisoning attacks in energy consumption prediction models of home appliances have been well assessed in this work. In future, we will work to develop more scalable poisoning attacks on different machine learning models and their defense mechanisms for the power industry.

**Author Contributions:** Conceptualization, Adnan Anwar, Ziaur Rahman and Mustain Billah; methodology, Adnan Anwar; software, Mustain Billah and Syed Md. Galib; validation, Adnan Anwar and Ziaur Rahman; formal analysis, Adnan Anwar and Ziaur Rahman; investigation, Adnan Anwar, Ziaur Rahman and Mustain Billah; resources, Adnan Anwar and Mustain Billah; data curation, Adnan Anwar and Mustain Billah; writing—original draft preparation, Mustain Billah; writing—review and editing, Adnan Anwar; visualization, Mustain Billah and Syed Md. Galib; supervision, Adnan Anwar; project administration, Adnan Anwar; funding acquisition, Adnan Anwar. All authors have read and agreed to the published version of the manuscript.

**Funding:** This research received no external funding

**Institutional Review Board Statement:** Not applicable

**Informed Consent Statement:** Not applicable

**Data Availability Statement:** The data presented in this study are openly available in the UCI Machine Learning Repository at "https://archive.ics.uci.edu/ml/datasets/Appliances+energy+prediction"

**Conflicts of Interest:** The authors declare no conflict of interest.